\documentclass[conference]{IEEEtran}
\IEEEoverridecommandlockouts

\usepackage{cite}
\usepackage{amsmath,amssymb,amsfonts}
\usepackage{algorithmic}
\usepackage{graphicx}
\usepackage{textcomp}
\usepackage{xcolor}
\usepackage{amsmath}
\usepackage{multirow}
\usepackage{makecell}
\usepackage{amsmath}
\usepackage{diagbox}

\def\BibTeX{{\rm B\kern-.05em{\sc i\kern-.025em b}\kern-.08em
    T\kern-.1667em\lower.7ex\hbox{E}\kern-.125emX}}
\begin{document}
\setlength{\skip\footins}{2pt}
\title{Singing Voice Conversion with Accompaniment Using Self-Supervised Representation-Based Melody Features
\\
}

\author{
\IEEEauthorblockN{
Wei Chen\textsuperscript{1},
Binzhu Sha\textsuperscript{1},
Jing Yang\textsuperscript{2},
Zhuo Wang\textsuperscript{2},
Fan Fan\textsuperscript{2},
Zhiyong Wu\textsuperscript{1}$^{\dagger}$\thanks{$^{\dagger}$ Corresponding author.}
}
\IEEEauthorblockA{\textsuperscript{1}Shenzhen International Graduate School, Tsinghua University, Shenzhen, China\\
}
\thanks{This work is supported by National Natural Science Foundation of China (62076144) and Shenzhen Science and Technology Program (WDZC20220816140515001, JCYJ20220818101014030).}
\IEEEauthorblockA{\textsuperscript{2}Huawei Technologies Co., Ltd., Shenzhen, China\\
\{chenw23, sbz22\}@mails.tsinghua.edu.cn, zywu@se.cuhk.edu.hk}
\{yangjing201, wangzhuo21, fanfan1\}@huawei.com,
\vspace{-18pt}}

\maketitle

\begin{abstract}
Melody preservation is crucial in singing voice conversion (SVC).
However, in many scenarios, audio is often accompanied with background music (BGM), which can cause audio distortion and interfere with the extraction of melody and other key features, significantly degrading SVC performance.
Previous methods have attempted to address this by using more robust neural network-based melody extractors, but their performance drops sharply in the presence of complex accompaniment.
Other approaches involve performing source separation before conversion, but this often introduces noticeable artifacts, leading to a significant drop in conversion quality and increasing the user’s operational costs.
To address these issues, we introduce a novel SVC method that uses self-supervised representation-based melody features to improve melody modeling accuracy in the presence of BGM.
In our experiments, we compare the effectiveness of different self-supervised learning (SSL) models for melody extraction and explore for the first time how SSL benefits the task of melody extraction.
The experimental results demonstrate that our proposed SVC model significantly outperforms existing baseline methods in terms of melody accuracy and shows higher similarity and naturalness in both subjective and objective evaluations across noisy and clean audio environments.
\end{abstract}
\vspace{-8pt}

\vspace{-10pt}
\section{Introduction}
\vspace{-4pt}
SVC is an emerging audio editing technology that aims to transform the timbre of an original singer into that of a target singer without altering other elements of a song (e.g., melody, lyrics).
Despite recent advancements in the field of speech conversion, SVC remains challenging due to the following reasons:
(1) Different musical styles lead to a greater variability in pitch, loudness, and articulation in singing, which makes modeling more complex;
(2) Human perception is highly sensitive to pitch errors in singing, as they are perceived as off-key and disrupt the melodic content;
(3) The scarcity of large-scale clean singing datasets impedes the generalization of SVC models.



Early SVC methods~\cite{kobayashi2014statistical,villavicencio2010applying,kobayashi2015statistical,toda2007one} address pitch-related challenges by leveraging parallel data to accurately recover the target pitch.
However, due to the scarcity of paired data, current SVC models focus on disentangling various aspects of singing, such as content, timbre, and pitch.
These models utilize singer-independent features, such as phonetic posteriorgrams (PPG)~\cite{li2021ppg,sun2016phonetic} and bottleneck features (BNFs)~\cite{ning2023vits} derived from Automatic Speech Recognition (ASR) models, as well as SSL representations~\cite{jayashankar2023self} trained on large amounts of unlabelled speech data, to directly generate target audio within the acoustic model.
Various deep generative models have been employed for decoding, including autoregressive models~\cite{zhang2020durian,takahashi2021hierarchical}, generative adversarial networks (GANs)~\cite{liu2021fastsvc}, variational autoencoders (VAE)~\cite{luo2020singing}, and diffusion models~\cite{liu2021diffsvc}.
However, these models can only perform voice conversion on clean vocal sequences, as they rely on non-robust methods (e.g. PYIN~\cite{mauch2014pyin}) for pitch extraction and directly use the foundamental frequency (F0) in SVC.

While many SVC models only deal with clean vocal data, 
singing is accompanied with BGM in real-world applications, which introduces distortions that interfere with the extraction of key acoustic features (e.g., pitch).
This interference significantly degrades SVC performance.
While some approaches~\cite{zhou2022hifi} employ more robust neural network-based pitch extractors, such as Crepe~\cite{kim2018crepe}, to mitigate this issue, these methods tend to perform poorly in the presence of higher noise levels or complex musical accompaniment.
Another approach to address this issue is to use music source separation algorithms to first remove the BGM from the audio before implementing SVC, which increases the barrier to user adoption.
Furthermore, these algorithms often introduce non-negligible artifacts, and when the processed audio is used as input to the SVC system, the overall conversion quality is still significantly reduced.

In this paper, we propose a novel any-to-one SVC model that can robustly generate clean singing voices directly from any source, even when the input contains complex background accompaniment.
The key to achieving this goal is to accurately extract melody features (including pitch information) from source audio. 
As demonstrated in ~\cite{wang2021towards,10219976}, SSL models show great potential in capturing rich acoustic and musical information.
Additionally,~\cite{hung22_interspeech} highlights the noise robustness of SSL models.
Based on this, we explore two distinct SSL models, HuBERT~\cite{hsu2021hubert} and WavLM~\cite{chen2022wavlm}, into our study.
By leveraging weighted embeddings from the outputs of each hidden layer of an SSL model, we ensure optimal extraction of melody features in noisy conditions.

This work makes the following contributions:
1) We propose a novel SVC method using self-supervised representation-based melody features to enhance the accuracy of melody modeling, even in the presence of BGM.
2) We compare the performance of various SSL models on melody extraction and are the first to investigate how SSL benefits the task of melody extraction.
3) The experimental results demonstrate that our SVC model significantly outperforms existing baseline methods in terms of melody accuracy, and shows higher similarity and naturalness in both subjective and objective evaluations under noisy and clean audio environments.

\begin{figure}[t]
  \centering
  \includegraphics[width=0.8\linewidth]{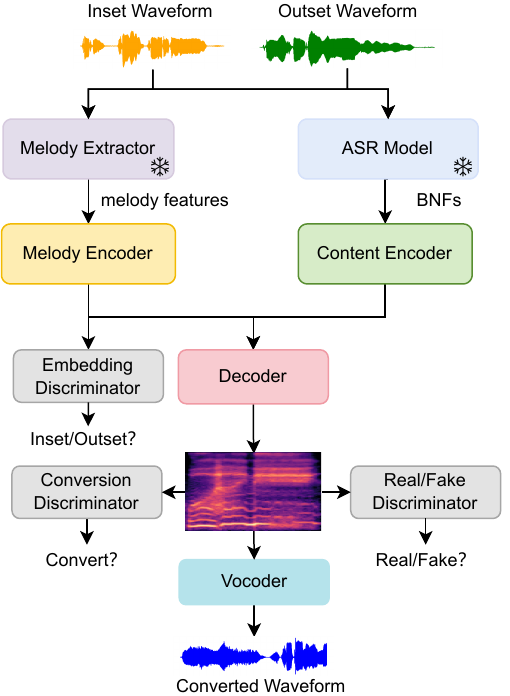}
  \caption{SVC framework. Snowflake represents the parameter that remains unchanged when training the SVC framework.}
  \label{fig:arc}
  \vspace{-10pt}
\end{figure}

\vspace{-6pt}
\section{Proposed Method}
\vspace{-4pt}
Fig.~\ref{fig:arc} illustrates an overview of our backbone Recognition-synthesis based SVC model~\cite{chen2024svc}.
We first train a melody extractor and an ASR model to extract melody features and BNFs containing content information.
These two features are then input into an encoder-decoder structured model.
We obtain high-quality Mel-spectrograms using three discriminators.
The Mel-spectrograms are further processed by a HiFi-GAN vocoder to obtain the converted audio.
In the following, we elaborate on the SVC framework and the melody extractor.
\vspace{-4pt}
\subsection{Singing Voice Conversion Framework}
\vspace{-4pt}
Referring to Fig.~\ref{fig:arc}, our model comprises five components: an ASR model, a melody extractor, an encoder-decoder architecture, discriminator groups and a vocoder.

The method employs the melody extractor and an ASR model to decouple the audio information and obtain melody features and content features (BNFs).
These features are separately fed into a melody encoder and a content encoder, both constructed with Feed Forward Transformer (FFT)~\cite{renfastspeech} blocks.
Notably, the melody encoder applies conditional instance normalization~\cite{dumoulin2016learned} after the FFT blocks to remove the timbre information of the source singer.
The decoder, based on FFT blocks, concatenates and decodes the input melody embeddings and content embeddings to generate Mel-spectrograms.

Additionally, the model uses adversarial learning strategies, training on both in-set and out-set data.
In-set data comprises songs sung by a single target singer, whereas out-set data includes songs from multiple singers.
Three types of discriminators are utilized to handle in-set and out-set data.
In order to synthesize high-quality Mel-spectrograms, the Real/Fake discriminator distinguishes whether the Mel-spectrograms are from the ground-truth audio or reconstructed from SVC model.
To enhance similarity with the target speaker, the Conversion discriminator judges if the Mel-spectrograms are converted from out-set singers or the target singer.
The Embedding discriminator determines whether melody embeddings originate from out-set songs, aiming to eliminate timbre information from the embeddings.

After the training of SVC framework, we fine-tune the pretrained HiFi-GAN vocoder~\cite{kong2020hifi} using ground-truth-aligned~\cite{shen2018natural} to address potential deficiencies in SVC model predictions.


\begin{figure}[t]
  \centering \hspace{10mm}
  \includegraphics[scale=0.9]{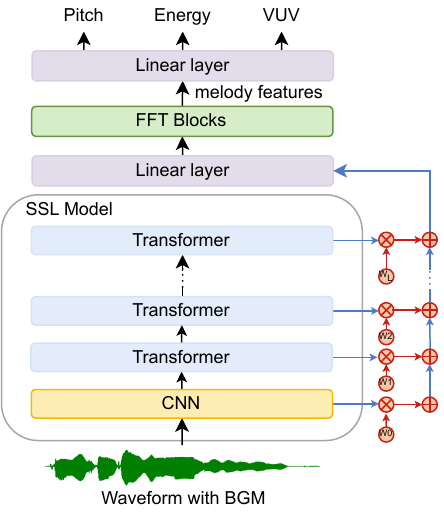}
  \caption{Melody extractor.}
  \label{fig:extract}
  \vspace{-10pt}
\end{figure}
\vspace{-6pt}
\subsection{Melody Extractor}
\vspace{-3pt}
Fig.~\ref{fig:extract} shows an overview of the melody extractor, which consists of three main components: an SSL model, weighted-sum, and FFT blocks.
We explore two distinct SSL models, HuBERT~\cite{hsu2021hubert} and WavLM~\cite{chen2022wavlm}, in our study.
To better capture melody information from the SSL representations, we apply weighted-sum\cite{chen22g_interspeech} to the hidden states of $L$ layer to generate the output representation $O = \left \{ o_{t}=\sum_{l=0}^{L} w_{l}\cdot h_{t}^{l} \right \} _{t=1}^{T}$, where $w_{l}$ is a learnable weight for the hidden state $h_{t}^{l}$ of the $l$-th layer, $T$ is the number of frames.
Incorporating FFT blocks, we further enhance the extraction of acoustic and musical information by feeding the output representation into these blocks to extract features for predicting pitch energy and voice/unvoice (VUV) flags. 
To improve the model robustness against various accompaniments, during training, the melody extractor is fed with singing data that includes BGM.
Additionally, we perform slight fine-tuning on them to fully leverage the potential of the SSL models.
Finally, the 256-dimensional output from the FFT blocks is chosen as the melody features input for the melody encoder. Note that the extracted feature contains more comprehensive melody information than pitch.  


\vspace{-5pt}
\section{Experimental Setup}
\vspace{-5pt}
\subsection{Dataset}
\vspace{-3pt}
To train the melody extractor, we utilize the MUSDB~\cite{rafii2017musdb18} and OpenSinger~\cite{huang2021multi} datasets.
The MUSDB dataset contains 10 hours of professional audio, composed of 150 songs with full supervision in stereo.
This dataset is divided into three subsets: 100, 35, and 15 songs for training, validation, and testing.
The OpenSinger dataset comprises 50 hours of high-quality Chinese vocals recorded in a professional studio, featuring 28 male and 48 female singers.
For the test set, we retain recordings from two male and two female singers, while the remaining recordings are randomly split into training and validation sets in a 9:1 ratio.

To train the SVC framework, we use the Opencpop~\cite{wang2022opencpop} corpus as in-set data, which includes 5.2 hours of recordings by professional female singers.
To enhance the diversity of the melodies, we augment the training data by varying the playback speed from 0.9x to 1.5x.
The OpenSinger dataset is used as out-of-domain data, with the same split methodology as that for training the melody extractor.


\vspace{-2pt}
\subsection{Training Conditions}\label{condition}
\vspace{-2pt}
During the training of the melody extractor, the ground-truth pitches are extracted from clean audio by taking the median of three methods: PYIN\footnote{https://github.com/librosa/librosa}, REAPER\footnote{https://github.com/google/REAPER}, and Parselmouth\footnote{https://github.com/YannickJadoul/Parselmouth}.
Root Mean Square Energy is computed from the STFT of the waveform with hop\_size$=$160.
The MUSDB dataset already involves BGM tracks and BGM-accompanied vocal tracks, but the OpenSinger dataset only involves clean vocal tracks. Therefore, we randomly select BGM from the MUSDB dataset and add BGM to the data in the OpenSinger dataset with a probability of 50\%. 
The AdamW optimizer with an initial learning rate of $2\times10^{-5}$ is employed.
we utilize two pretrained SSL models: HuBERT and WavLM, to demonstrate the generalizability of our method.
Both models are trained on the Librispeech~\cite{panayotov2015librispeech} dataset.
To evaluate the effectiveness of fine-tuning the SSL model, weighted-sum, and FFT blocks, we train the melody extractor under the following conditions: \textit{raw single}, \textit{raw weighted-sum}, \textit{single w/o FFT}, \textit{weighted-sum w/o FFT}, \textit{single w/ FFT}, \textit{weighted-sum w/ FFT}, and \textit{proposed}.
\textit{Raw single} means directly using the last-layer output from the pretrained SSL model as the melody features.
\textit{Raw weighted-sum} means using the weighted-sum of each layer output from the pretrained SSL model as the melody features.
\textit{Single w/o FFT} means using the last layer output from the SSL model and training the melody extractor without FFT blocks.
\textit{Weighted-sum w/o FFT} means weighted-sum each layer output from the SSL model and training the melody extractor without FFT blocks.
\textit{Single w/ FFT} means using the last layer output from the pretrained SSL model and training the melody extractor with FFT blocks.
\textit{Weighted-sum w/ FFT} means weighted-sum each layer output from the pretrained SSL model and training the melody extractor with FFT blocks.
Finally, \textit{proposed} means weighted-sum each layer output from the SSL model and training the melody extractor with FFT blocks.
Note that when the SSL model is involved in training, we only train it for 5k steps and then freeze the SSL model's parameters to mitigate the risk of catastrophic forgetting~\cite{french1999catastrophic}.


Referring to Fig.~\ref{fig:arc}, when training the SVC, an ASR model is initially trained using Connectionist Temporal Classification loss, following the configuration described in ~\cite{zhao2022disentangling}, and utilizing the WeNet~\cite{yao2021wenet} implementation\footnote{https://github.com/wenet-e2e/wenet}.
256-dimensional BNFs are extracted from the ASR model.
All audio files are down-sampled to 16 kHz, and Mel-spectrograms are generated using STFT with a 50 ms frame size, 10 ms frame hop, and a Hanning window function.
To compare with state-of-the-art (SoTA) methods, we directly input pitch, content embeddings and energy into the decoder to train the SVC model.
The pitch is obtained using two methods: (1) the method mentioned in~\ref{condition}, referred to as \textit{"Original Pitch\&Energy"}, and (2) a noise-robust neural F0 extractor called Crepe~\cite{kim2018crepe}, referred to as \textit{"Crepe"}.
Additionally, we employ the \textit{"Separated+crepe"} approach, in which we first separated songs into clean vocal tracks and BGMs using Demucs~\cite{rouard2023hybrid}, and then extract pitch and energy from the clean vocal tracks using the Crepe model. 
Both the baseline and proposed systems are evaluated under identical conditions unless stated otherwise.
Audio samples are available on our demo page\footnote{https://ccchange2024.github.io/SVC-with-Accompaniment}.

\vspace{-4pt}
\section{Result}
\vspace{-4pt}
\subsection{Melody Features Evaluation}
We first evaluate the performance of melody features extraction under different conditions for SVC.
The source audio used for testing contains unseen BGM from the test set, with signal-to-noise ratios (SNR) of 0, 5, 10, and 15, each accounting for 25\% of the dataset.
To assess the quality of the converted audio, we calculate the F0 Root Mean Square Error (F0RMSE) and the correlation coefficient (F0CORR)~\cite{huang2023singing}.
F0RMSE measures the naturalness of the converted audio, and the F0 sequence is normalized using min-max scaling before processing.
The RMSE between the converted and source waveforms is computed in the absence of BGM.
F0CORR evaluates pitch accuracy by aligning F0 contours of the original and synthesized audio using dynamic time warping and then computing the Pearson correlation coefficient.
This metric highlights the model’s ability to preserve pitch, which is critical for achieving natural and expressive audio.

As shown in Table.~\ref{tab:my_label}, the comparison between the conditions with and without weighted-sum demonstrates the superiority of the weighted-sum approach.
Most fine-tuning methods show performance improvements compared to the pretrained methods.
Notably, the pretrained \textit{weighted-sum w/ FFT} outperforms the \textit{single w/o FFT} during fine-tuning, benefiting from more comprehensive melody information extraction in the SSL model.
The proposed method achieves the best performance across two different SSL models, and since WavLM exhibits superior results, we choose WavLM for the proposed method and for further comparisons with other melody extraction models.

\begin{table}[t]
    \caption{Results of different melody input.}
    \label{tab:my_label}
    \centering \hspace{-3mm}
    \scalebox{0.95}{
    \begin{tabular}{|c|c|c|c|c|c|} \hline  
         \multirow{2}{*}{Method}&  \multicolumn{3}{c|}{Condition} & \multirow{2}{*}{F0RMSE$\downarrow$}& \multirow{2}{*}{F0CORR$\uparrow$}\\ \cline{2-4}
         &  FT&  WS&  FFT&  & \\ \hline  
         H-single&  &  &  &  0.360& 0.741\\ \hline  
         H-weighted-sum&  &  $\checkmark $&  &  0.239& 0.745\\ \hline  
         H-single w/o FFT&  $\checkmark $&  &  &  0.246& 0.799
\\ \hline  
         H-weighted-sum w/o FFT&  $\checkmark $&  $\checkmark $&  &  0.228& 0.867
\\ \hline  
         H-single w/ FFT&  &  &  $\checkmark $&  0.287& 0.795
\\ \hline  
         H-weighted-sum w/ FFT&  &  $\checkmark $&  $\checkmark $&  0.234& 0.829
\\ \hline  
         H-proposed&  $\checkmark $&  $\checkmark $&  $\checkmark $&  \textbf{0.202}& \textbf{0.899}
\\ \hline  \hline
         W-single&  &  &  

&  0.320& 0.740
\\ \hline  
 W-weighted-sum& & $\checkmark $& 

& 0.289&0.755
\\ \hline  
 W-single w/o FFT& $\checkmark $& & 

& 0.230&0.887
\\ \hline  
 W-weighted-sum w/o FFT& $\checkmark $& $\checkmark $& 

& 0.194&0.932
\\ \hline  
 W-single w/ FFT& & & $\checkmark $
& 0.260&0.811
\\ \hline  
 W-weighted-sum w/ FFT& & $\checkmark $& $\checkmark $
& 0.201&0.914
\\ \hline  
 W-proposed& $\checkmark $& $\checkmark $& $\checkmark $& \textbf{0.176}&\textbf{0.950}
\\ \hline 
 \multicolumn{6}{l}{H represent using Hubert, W represent using WavLM.}\\
 \multicolumn{6}{l}{FT, WS, and FFT represent fine-tuning, weighted-sum part and FFT blocks.} \\
 \multicolumn{6}{l}{Values in bold show the best scores in each noise condition.}\\
    \end{tabular}
    }
    
    \vspace{-20pt}
\end{table}

\vspace{-4pt}
\subsection{Comparison with State-of-the-art Methods}
\vspace{-4pt}
We conduct both objective and subjective evaluations of various SoTA methods under clean and with BGM conditions.
The objective experiments utilize the F0RMSE and F0CORR metrics.
For the subjective evaluation, we employ a 5-point Mean Opinion Score (MOS) test and invite 15 volunteers with extensive knowledge of music theory to participate.
Notably, the subjective test includes 12 audio samples for both clean and BGM conditions.
The samples with BGM are evenly distributed across SNR of 0, 5, 10, and 15, with three audio samples for each SNR level.

As shown in Table.~\ref{tab:compare}, under clean conditions, the objective metrics show similar performance across different methods.
However, in the subjective tests, our proposed method demonstrates superior performance in terms of both similarity and naturalness, even achieving scores close to those of the source audio.
The \textit{Separated+Crepe} method performs relatively poorly due to the loss of some vocal information in the separated signal, leading to instability in SVC.
In BGM conditions, all methods show a decline in performance as the SNR decreased. Nonetheless, our proposed method consistently achieves the best results, both objectively and subjectively.

\begin{table}[t]
    \caption{Comparison with state-of-the-art methods}
    \label{tab:compare}
    \centering
    \scalebox{0.90}{
    \begin{tabular}{|>{\centering\arraybackslash}p{0.11\linewidth}|>{\centering\arraybackslash}p{0.15\linewidth}|>{\centering\arraybackslash}p{0.12\linewidth}|>{\centering\arraybackslash}p{0.12\linewidth}|c|c|} \hline 
         Method&  SNR Level&  F0RMSE$\downarrow$&  F0CORR$\uparrow$&  NMOS& SMOS\\ \hline
         \multirow{5}{*}{\makecell{Original\\Pitch\&\\Energy}}&  clean&  0.165&  0.965
&  3.38±0.14& 3.58±0.13\\ \cline{2-6}
         &  15dB&  0.261&  0.903
&  \multirow{4}{*}{2.30±0.13}& \multirow{4}{*}{2.88±0.14}\\ 
         &  10dB&  0.306&  0.889
&  & \\ 
         &  5dB&  0.325&  0.818
&  & \\ 
         &  0dB&  0.353&  0.779
&  & \\ \hline
         \multirow{5}{*}{Crepe}&  clean&  \textbf{0.164}&  \textbf{0.967}
&  3.07±0.16& 3.52±0.13\\ \cline{2-6}
         &  15dB&  0.208&  0.915
&  \multirow{4}{*}{2.40±0.14}& \multirow{4}{*}{3.04±0.14}\\ 
         &  10dB&  0.261&  0.895
&  & \\ 
         &  5dB&  0.301&  0.856
&  & 
\\ 
 & 0dB& 0.297& 0.815
& &
\\ \hline
 \multirow{5}{*}{\makecell{Separated\\+\\Crepe}}& clean& 0.171& 0.950
& 3.03±0.16&
3.42±0.12\\ \cline{2-6}
 & 15dB& 0.176& 0.946
& \multirow{4}{*}{2.57±0.16}&
\multirow{4}{*}{3.07±0.14}\\ 
 & 10dB& 0.178& 0.943
& &
\\ 
 & 5dB& \textbf{0.182}& 0.942
& &

\\ 
 & 0dB& 0.202& 0.928
& &\\ \hline
 \multirow{5}{*}{\makecell{proposed\\with\\WavLM}}& clean& 0.169& 0.960
& \textbf{3.54±0.14}&\textbf{3.71±0.14}\\ \cline{2-6}
 & 15dB& \textbf{0.174}& \textbf{0.955}
& \multirow{4}{*}{\textbf{3.49±0.14}}&\multirow{4}{*}{\textbf{3.56±0.13}}\\ 
 & 10dB& \textbf{0.175}& \textbf{0.949}
& &\\ 
 & 5dB& 0.183& \textbf{0.943}
& &\\ 
 & 0dB& \textbf{0.199}& \textbf{0.935}
& &\\\hline
 source& clean &\diagbox[width=42pt,height=10pt]{}{} & \diagbox[width=42pt,height=10pt]{}{}& 4.55±0.11&\diagbox[width=50pt,height=10pt]{}{}\\ \hline 
 \multicolumn{6}{l}{MOS results are reported with 95\% confidence intervals.}\\
    \end{tabular}
    }
    
    \vspace{-10pt}
\end{table}

\begin{figure}[t]
  \centering
  \includegraphics[width=\linewidth]{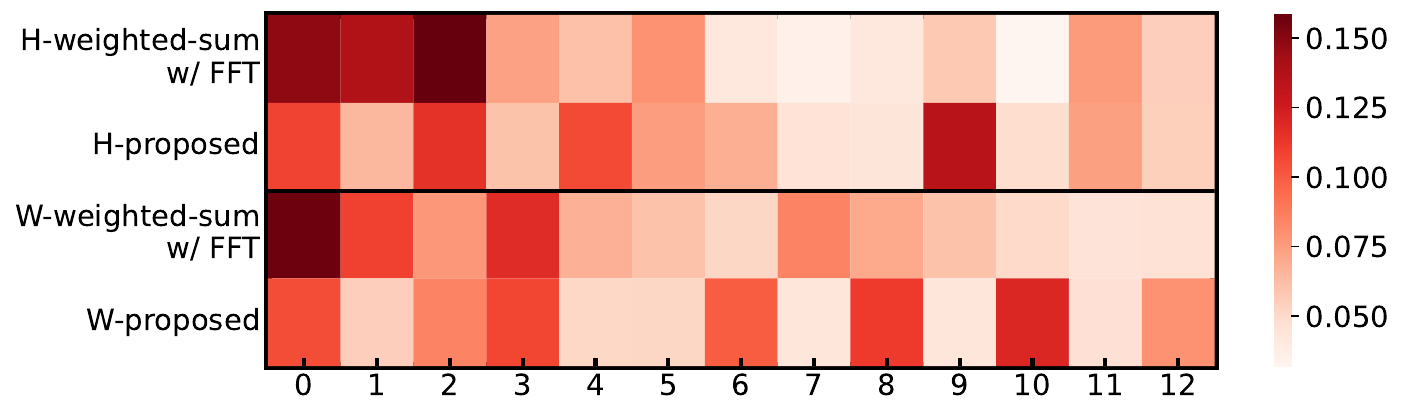}
  \caption{Visualized weight that extract representations from SSL models.}
  \label{fig:weight}
  \vspace{-10pt}
\end{figure}
\vspace{-4pt}
\subsection{Contribution Analysis for Weighted-sum}
\vspace{-4pt}
We visualize the weights of weighted-sum part under both pretraining and fine-tuning conditions to illustrate how each layer of the pretrained model contributes to the final melody extraction performance.
From Fig.~\ref{fig:weight}, we observe that in the pretrained model, the main contributions come from the output of the CNN feature extractor and the lower layers of encoder for all the pretrained models.
It indicates that during SSL, only the shallow layers of the pretrained model capture melody-related information.
In contrast, during fine-tuning, we update both the downstream module parameters and the pretrained parameters.
By unleashing the full potential of the pretrained model, the higher Transformer-based encoder layers also learn to model melody information using the melody extraction objective.
This allows the higher layers to make a greater contribution to the final prediction compared to the pretrain stage, ultimately leading to improved melody extraction performance.

\vspace{-6pt}
\section{Conclusion}
\vspace{-6pt}
We have proposed a novel any-to-one SVC model that can generate clean singing voices even when the input includes BGM.
Both objective and subjective evaluations demonstrate that our method exhibits greater robustness to BGM compared to other methods.
Additionally, we explore how self-supervised models learn melody information and found that only the shallow layers of the pretrained model capture melody-related information.
After fine-tuning, the higher Transformer-based encoder layers are also able to model melody information using the melody extraction training objective and have better melody extraction performance.
Future work will focus on applying our proposed method to more complex noisy environments and extending it to any-to-many SVC tasks.


\bibliographystyle{IEEEtran}

\bibliography{mybib}

\end{document}